\documentclass[nologo,url,11pt,a4paper]{ETHpaper}

\usepackage{fancyhdr,epsfig,url,hyperref,lastpage}
\usepackage[dvips]{color}
\usepackage[square,sort&compress]{natbib}

\newcommand{\be}{\begin{equation}}
\newcommand{\ee}{\end{equation}}

\newcommand{\mean}[1]{\left\langle #1 \right\rangle}
\newcommand{\abs}[1]{\left| #1 \right|}

\title{A Model of a Trust-based Recommendation \\ System on a Social Network}
\titlealternative{A Model of a Trust-based Recommendation System on a Social Network}

\author{Frank E. Walter, Stefano Battiston, and Frank Schweitzer}

\address{Chair of Systems Design, ETH  Zurich, Kreuzplatz 5, 8032 Zurich, Switzerland \\
  \url{fewalter@ethz.ch}, \url{sbattiston@ethz.ch}, \url{fschweitzer@ethz.ch}}

\reference{Revised Manuscript for JAAMAS (September 07, 2007).}

\www{\url{http://www.sg.ethz.ch}}

\begin{document}

\maketitle

\begin{abstract}

  In this paper, we present a model of a trust-based recommendation
  system on a social network. The idea of the model is that agents use
  their social network to reach information and their trust relationships
  to filter it. We investigate how the dynamics of trust among agents
  affect the performance of the system by comparing it to a
  frequency-based recommendation system. Furthermore, we identify the
  impact of network density, preference heterogeneity among agents, and
  knowledge sparseness to be crucial factors for the performance of the
  system. The system self-organises in a state with performance near to
  the optimum; the performance on the global level is an emergent
  property of the system, achieved without explicit coordination from the
  local interactions of agents.

\end{abstract}

\textbf{Keywords:} Recommender System, Trust, Social Network

\section{Introduction and Motivation}

In recent years, the Internet has become of greater and greater importance in
everyone's life. People use their computers for communication with others, to
buy and sell products on-line, to search for information, and to carry out many
more tasks. The Internet has become a social network, ``linking people,
organisations, and knowledge'' \citep{wellman01} and it has taken the role of a
platform on which people pursue an increasing amount of activities that they
have usually only done in the real-world.

This development confronts people with an \textit{information overload}:
they are facing too much data to be able to effectively filter out the
pieces of information that are most appropriate for them. The exponential
growth of the Internet \citep{huberman99} implies that the amount of
information accessible to people grows at a tremendous rate.
Historically, people have -- in various situations -- already had to cope
with information overload and they have intuitively applied a number of
\textit{social mechanisms} that help them deal with such situations.
However, many of these, including the notion of trust, do not yet have an appropriate
\textit{digital mapping} \citep{marsh94}. Finding suitable representations
for such concepts is a topic of on-going research
\citep{castelfranchi01,mui02a,sabater05,grandison00,marsh94,abdul-rahman00}.

The problem of information overload has been in the focus of recent research
in computer science and a number of solutions have been suggested. The use of
search engines \citep{brin98} is one approach, but so far, they lack
personalisation and usually return the same result for everyone, even though
any two people may have vastly different profiles and thus be interested in
different aspects of the search results.  A different proposed approach are
recommendation systems \citep{montaner02b,montaner03a,massa04a,sarwar00}.

In the following, we present a model of a trust-based recommendation
system which, in an automated and distributed fashion, filters
information for agents based on the agents' social network and trust
relationships \citep{montaner02b,golbeck06a,guha04}.

Trust is a topic which has recently been attracting research from many
fields, including, but not limited to, computer science, cognitive
sciences, sociology, economics, and psychology. As a result of this,
there exists a plethora of definitions of trust, some similar to each
other, some different from each other. In the context of our model, trust
can be defined as the \textit{expectancy of an agent to be able to rely on some
other agent's recommendations}.

There are many areas of application in which such systems, or similar
ones, are applicable: some obvious examples would be the facilities to
share opinions and/or ratings that many shopping or auctioning web sites
offer, but the same principles of combining social networks and trust
relationships can be applied in other domains as well: for example, in
the scientific community, in form of a recommendation system for journal,
conference, and workshop contributions.

The model that we are going to present enables a quantitative study of
the problem and also provides a sketch for a solution in terms of a real
Internet application/web service.  The idea at the core of the model is
that agents

\begin{itemize}
\item leverage their social network to \textit{reach information}; and
\item make use of trust relationships to \textit{filter information}.
\end{itemize}

In the following, we describe the model and the results obtained by
simulating the model with multi-agent simulations. To some extent, it is
also possible to make analytical predictions of the performance of the
system as a function of the preferences of the agents and the structure
of the social network.

The remainder of the paper is organised as follows: in the following
section, we put our work into the context of the related work. Then, we
present our model of a trust-based recommendation system on a social
network. This is followed by an analysis of the results from computer
simulations and analytical approximations of the model. Subsequently, we
illustrate a number of possible extensions and conclude with a summary of
the work.

\section{Related Work}

Recent research in computer science has dealt with recommendation systems
\citep{sarwar00,montaner03a}.  Such systems mostly fall into two classes:
content-based methods suggest items by matching agent profiles with
characteristics of products and services, while collaborative filtering
methods measure the similarity of preferences between agents and
recommend what similar agents have already chosen \citep{sarwar00}. Often,
recommendation systems are centralised and, moreover, they are offered by
entities which are not independent of the products or services that they
provide recommendations on -- often, this constitutes a bias or
conflict-of-interest.

Trust is a topic which is of ubiquitous importance to people. This is why
it has been studied in many disciplines, among them computer science,
cognitive sciences, sociology, economics, and psychology
\citep{abdul-rahman00,castelfranchi01,grandison00,marsh94,sabater05}. In computer
science, trust was initially seen as a method to enhance security systems
\citep{grandison00}: cryptography allows to ensure the authenticity,
confidentiality, and integrity, of the communication between two parties
Alice and Bob, but it does not allow Alice to judge how trustworthy Bob
is, and vice versa. In such contexts, trust has often been formalised
with logical models \citep{huang06,josang05}. For a more detailed
overview of trust in the literature, please refer to \citep{sabater05}.

Additionally, the diffusion of information technologies in business and
social activities results in intricate networks of electronic
relationships. In particular, many economic activities via electronic
transactions require the presence of or benefit from a system of trust
and distrust in order to ensure the fulfilment of contracts
\citep{marsh94, sabater05}.  However, trust plays a crucial role not only
by supporting the security of contracts between agents, but also because
agents rely on the expertise of other trusted agents in their
decision-making.

Along these lines, some recent works have suggested to combine
distributed recommendation systems with trust and reputation mechanisms
\citep{montaner02b,massa04a,golbeck06b}. Because of the fact
that both building expertise and testing items available on the market
are costly activities, individuals in the real world attempt to reduce
such costs through the use of their social/professional networks.

Such complex networks, in particular their structure and function, are
the subject of an extensive and growing body of research across
disciplines \citep{newman03a}.  Social networks have received special
attention \citep{battiston04} and it has been shown that their
structure plays an important role in decision making processes
\citep{garlaschelli03, battiston03a, battiston03b}.

With respect to existing models of trust-based recommender systems
operating on social networks in the literature
\citep{golbeck06b,guha04,montaner02b}, the contributions of our work are
the following: we provide analytical results for the performance of the
system within a range of network density, preference heterogeneity among
agents, and knowledge sparseness. We also report on extensive multi-agent
simulations supporting our predictions. The notion of trust that we use
is quite general because it relies on the utility of an agent from
interacting with other agents. Thus, it can be extended to represent more
than just the similarity of preferences between two agents
\citep{golbeck06c,ziegler06}. With respect to \citep{montaner02b},
besides the above, our model includes a mechanism for propagation of
trust along paths in the social network. Finally, we provide a framework
which allows the study of two crucial aspects, evolution and robustness,
both from an analytical point-of-view, but also by multi-agent
simulations; in this respect, the framework could be validated against
empirical data along the lines of \citep{massa04a}.

\section{Model Description}

The model deals with agents which have to decide for a particular item
that they do not yet know based on recommendations of other agents. When
facing the purchase of an item, agents query their neighbourhood for
recommendations on the item to purchase. Neighbours in turn pass on a
query to their neighbours in case that they cannot provide a reply
themselves. In this way, the network replies to a query of an individual
by offering a set of recommendations. One way to deal with these
recommendations would be to choose the most frequently recommended item.
However, because of the heterogeneity of preferences of agents, this may
not be the most efficient strategy in terms of utility. Thus, we explore
means to incorporate knowledge of trustworthiness of recommendations into
the system. In the following, we investigate, by means of analytical
calculations and computer simulations, under which conditions and to what
extent the presence of a trust system enhances the performance of a
recommendation system on a social network.

\subsection{Agents, Objects, and Profiles}

We consider a set $S_A$ of $N_A$ agents $a_1, a_2, a_3, ..., a_{N_A}$.
The idea is that the agents are connected in a \textit{social network}
such as, for example, of people and their friends
\citep{watts98,barabasi99,newman03a} that are recommending books to each
other.  Hence, each agent has a set of links to a number of other agents
(which we call its neighbours). These links are not necessarily
symmetric, i.e. the graph is directed. In reality, social networks
between agents evolve over time; in other words, relationships form,
sustain, and also break up. In this paper, we mainly focus on a static
network while dynamic networks will be investigated more thoroughly in
further work. At this stage, we assume the network to be described by a
random graph \citep{erdos59,bollobas85} -- the usual choice in absence of
knowledge of the real structure of the modeled social network. We are
aware that random graphs are not always a good approximation of real
networks. Thus, for further analysis of the model, it will be appropriate
to experiment with several different topologies as discussed in
\citep{amaral00}.

Furthermore, there exists a set $S_O$ of $N_O$ objects, denoted $o_1,
o_2, o_3, ..., o_{N_O}$. These objects represent items, agents, products,
buyers, sellers, etc. -- anything that may be subject to the
recommendations -- i.e. books as in the running example. We further
assume that objects are put into one or more of $N_C$ categories from
$S_C$, denoted $c_1, c_2, ..., c_{N_C}$, where these categories are
defined by the system and cannot be modified (i.e.  added, removed, or
redefined) by the agents. In the scenario where the recommendation system
is on books, categories could be `epicurean philosophy', `Swiss
folklore', or `medieval archery'.  We denote the fact that an object
$o_i$ is in category $c_j$ by stating $o_i \in c_j$.

\begin{figure}[htbp]
  \centering
  \includegraphics[width=0.4\textwidth]{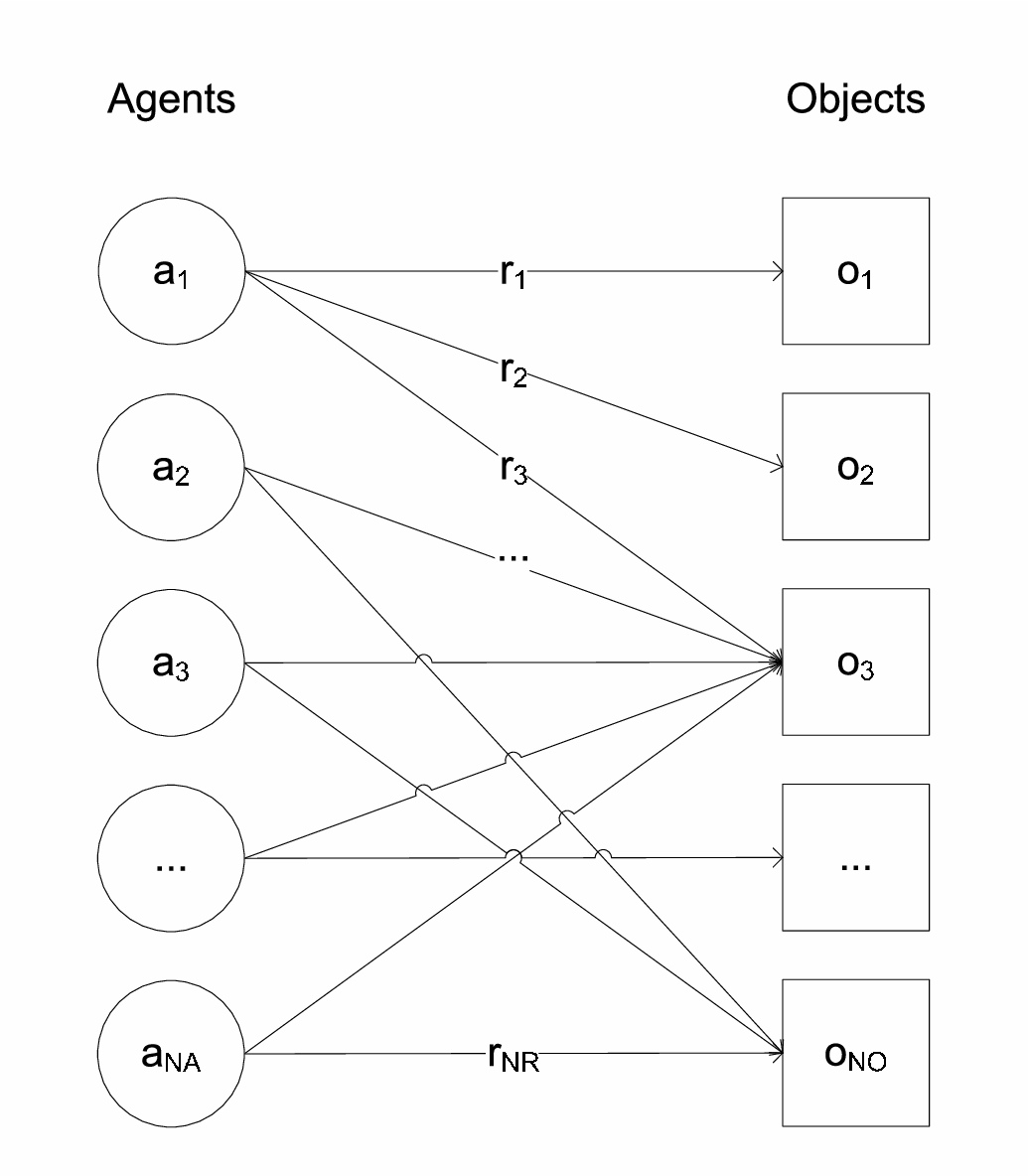}
  \vspace{-5mm}
  \caption[Agents rating Objects]{Agents rating Objects: this is a bipartite
  graph with the agents on the left hand side and the objects on the right
    hand side, the ratings being the connections. The set of all possible
    ratings of an agent constitutes its respective profile.}
  \label{fig:agents-objects}
\end{figure}

Each agent $a_i$ is associated to one certain preference profile which is
one of $N_P$ preference profiles in the system, where $S_P = \{p_1, p_2,
p_3, ..., p_{N_P}\}$. In the following, we will use the terms `preference
profile', `profile', and 'preferences' interchangeably. Such a profile
$p_i$ is a mapping which associates to each object $o_j \in S_O$ a
particular corresponding rating $r_j \in [-1,1]$, $p_i : S_O \rightarrow
[-1,1]$.  This is illustrated in Figure \ref{fig:agents-objects}. In the
current version of the model, we only consider discrete ratings where
$-1$ signifies an agents' dislike of an object, $1$ signifies an agents'
favour towards an object. In a future version of the model, this
assumption can be relaxed; we chose to initially focus on a discrete
rating scheme because most of the ones found on the Internet are of such
type. We assume that agents only have knowledge in selected categories
and, in particular, they do only know their own ratings on objects of
other categories subsequent to having used these objects.  Thus, each
agent is and remains an expert only on a set of initially assigned
selected categories.

\subsection{Trust Relationships}

\begin{figure}[htb]
  \centering
  \includegraphics[width=0.50\textwidth]{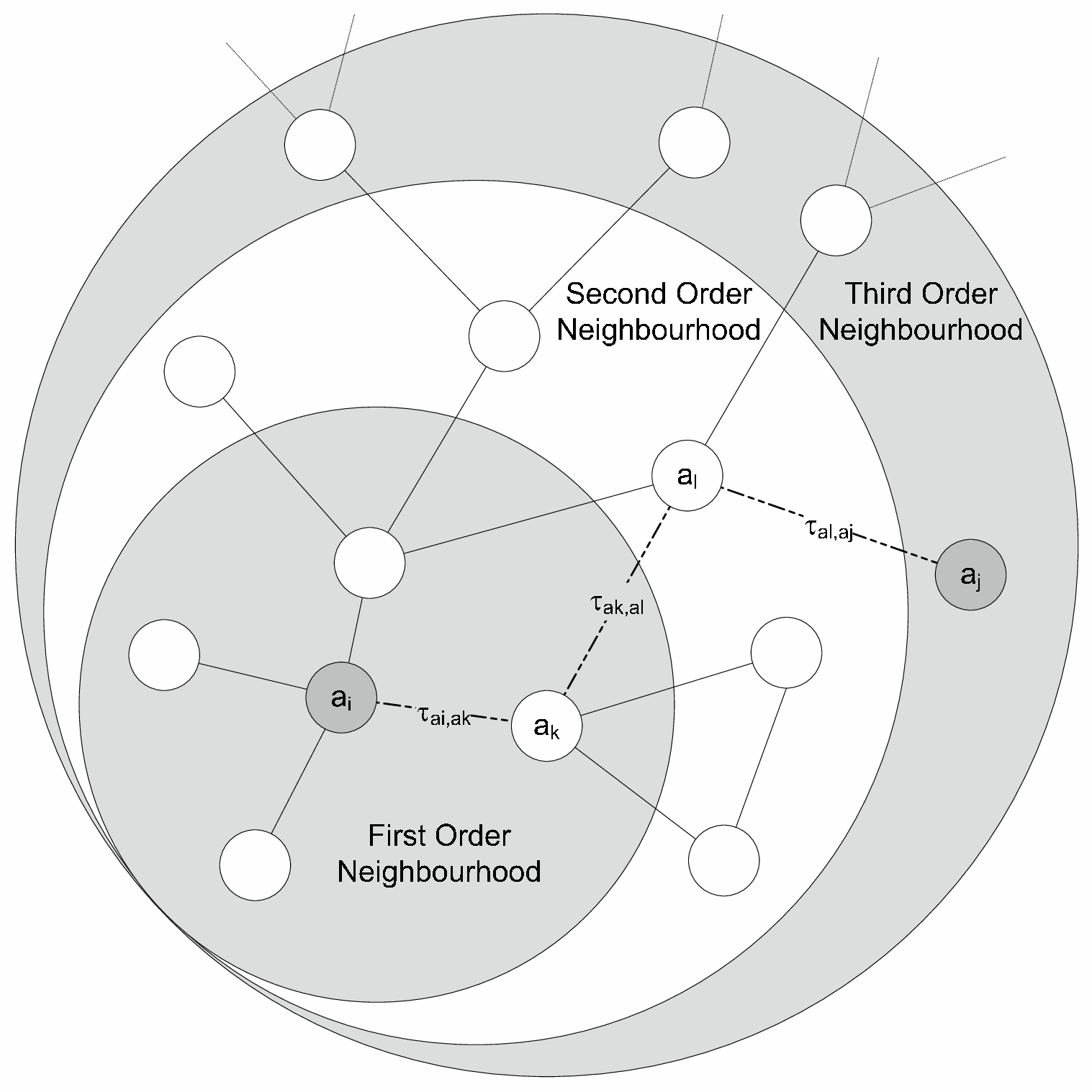}
  \vspace{-5mm}
  \caption[Social Network of Agents and their Trust Relationships]{Social
    Network of Agents and their Trust Relationships: a section of the social
    network around agent $a_i$, indicating a chain of trust relationships to
    agent $a_j$ and ordering the neighbours according to their distance in
    hops ('orders of neighbourhood').}
  \label{fig:social-network-trust-relationships}
\end{figure}

In this model, we also consider trust relationships between agents: each
agent $a_i$ keeps track of a trust value $T_{a_i,a_j} \in [0,1]$ to
each of its neighbour agents $a_j$.  These values are initialised to
$T_{a_i,a_j} = 0.5$. It is important to stress that trust
relationships only exist between neighbours in the social network; if two
agents are not directly connected, they also cannot possibly have a trust
relationship with each other. However, two such agents may indirectly be
connected to each other through a path in the network. For example, agent
$a_i$ could be connected to agent $a_j$ through agents $a_k$ and $a_l$,
should $a_k$ and $a_l$, $a_i$ and $a_k$, as well as $a_l$ and $a_j$ be
neighbours. We can then compute a trust value along the path
$\textrm{path}(a_i,a_j)$ from $a_i$ to $a_j$ -- in the example,
$\textrm{path}(a_i,a_j)=\{(a_i,a_k),(a_k,a_l),(a_l,a_j)\}$ -- as follows:

\begin{eqnarray}
  T_{a_i, ..., a_j}=\prod_{(a_k, a_l) \in \textrm{path}(a_i,a_j)}{T_{a_k, a_l}}
  \label{trust_relationships}
\end{eqnarray}

i.e. the trust value along a path is the product of the trust values of
the links on that path. Of course, there may be more than one path
between two agents; in such cases, each path has its own trust value.
Figure \ref{fig:social-network-trust-relationships} illustrates a part of
such a social network of agents and a chain of trust relationships
between two agents.

Note that this implies the assumption that trust is able to propagate
through the network. In other words, we take the position that ``if $i$
trusts $j$, and $j$ trusts $k$, then $i$ should have a somewhat more
positive view of $k$ based on this knowledge'' \citep{guha04}.

Trust transitivity is a condition for trust propagation and there have
been fierce discussions in the literature whether or not trust is
transitive.  From the perspective of network security (where transitivity
would, for example, imply accepting a key with no further verification
based on trust) or formal logics (where transitivity would, for example,
imply updating a belief store with incorrect, impossible, or inconsistent
statements) it may make sense to assume that trust is not transitive
\citep{josang05, huang06, christianson97}. Others attribute trust some
degree of transitivity \citep{gray03,guha04}.  Furthermore, it has been shown
empirically \citep{guha04,golbeck06a,golbeck06b} that in scenarios similar
to ours, it can be assumed that ``trust may propagate (with appropriate
discounting) through the relationship network'' \citep{guha04}. In our
model, discounting takes place by multiplying trust values along paths.

It is important to remark that we do not allow recommendations by other
agents to influence the preferences of an agent on items. Rather,
recommendations are suggestions. An agent merely selects one and then,
based on its experience, draws its own consequences, regardless of the
recommendation.

\subsection{Temporal Structure, Search for Recommendations}

The model assumes a \textit{discrete linear bounded model of time}. In
essence, there are two possible types of search for a recommendation:

\begin{enumerate}
\item \textit{Ranking within a category (RWC)}: agents query for a
  particular category and search recommendations for several objects in
  this category in order to decide for one of the recommended objects in
  the response from the network -- typically the best one.
\item \textit{Specific rating for an object (SRO)}: agents query for a
  particular object and search recommendations on this very object in
  order to decide for or against using it, based on the response from the
  network.
\end{enumerate}

Both variants are possible within the framework of our model: in fact, we
use SRO to establish RWC.

At each time step $t$, each agent $a_i$ (in random order) selects a
category $c_j$ (again, in random order, with the constraint that the
agent is not an expert on the category) and searches for recommendations
on the network. The protocol for agent's search proceeds as follows:

\begin{enumerate}
\item Agent $a_i$ prepares a $\textrm{query}(a_i, c_j)$ for category
  $c_j$ and then transmits it to its neighbours.

\item Each neighbour $a_k$ receives $\textrm{query}(a_i, c_j)$ and either
  \begin{enumerate}
  \item returns a $\textrm{response}(a_k, a_i, (o_j, r_j),
    T_{a_i,...,a_k})$, if it knows a rating $r_j$ for a particular object
    $o_j$ in $c_j$ that it can recommend, i.e. if $p_k(o_j)=r_j>0$ (only
    positive ratings are communicated, we do not consider negative ones
    at this stage for sake of simplicity in the decision making process)

  \item or, passes $\textrm{query}(a_i, c_j)$ on to its own neighbours if it
    does not know a rating $r_j$ for the particular category $c_j$.

  \end{enumerate}
\end{enumerate}

Notice that an agent $a_k$ knows the rating $r_j$ for a particular object
$o_j$ only if it has either experienced it or if it is an expert on the
category $c_j$ of the object. Furthermore, each agent along the path
computes only a part of the product $T_{a_i,...,a_k}$ -- i.e. on the path
path($a_1$, $a_2$, $a_3$, $a_4$), agent $a_3$ would pass $T_{a_3,a_4}$ to
$a_2$ and then, $a_2$ would compute
$T_{a_2,a_3,a_4}=T_{a_2,a_3}T_{a_3,a_4}$ and pass it to agent $a_1$ who
then can compute $T_{a_1,a_2,a_3,a_4}=T_{a_1,a_2}T_{a_2,a_3,a_4}$.

It is assumed that agents keep track of the queries they have seen. Now
there are two strategies to guarantee that the algorithm terminates: either,

\begin{itemize}
\item agents do not process queries that they have already seen again
  (``incomplete search'', IS); or,
\item agents pass on queries only once, but, if they have an appropriate
  recommendation, can return responses more than once (``complete
  search'', CS).
\end{itemize}

In essence, both are a form of \textit{breadth-first search} on the
social network of agents, but with different properties: the former
returns, for each possible recommendation, only one possible path in the
network from the querying to the responding agent; the latter, however,
returns, for each possible recommendation, each of the possible paths in
the network from the querying to the responding agent.

As we will see later, this is a crucial difference for the decision
making of agents. For a given recommendation, there might be several
paths between the querying and the responding agent. The IS returns a
recommendation along one of these paths, while the CS returns a set of
recommendations along all possible paths.  Some paths between two agents
have high trust, some have low trust. The IS may return a recommendation
along a low-trust path even though there exists a high-trust path, thus
providing an agent with insufficient information for proper decision
making. Of course, there is also a pitfall with the CS -- it is
computationally much more expensive. In the literature, this issue of
potentially having multipe paths for a recommendation has been discussed
\citep{gray03}, and we will come back to it when discussing the decision
making of the agents.

\subsection{Decision Making}

As a result of a query, each agent $a_i$ possesses a set of responses
from other agents $a_k$. It now faces the issue of making a decision for
a particular object. The agent needs to decide, based on the
recommendations in the response, what would be the appropriate choice of
all the objects recommended. In the following, we denote
$\textrm{query}(a_i, o_j) = Q$ and a $\textrm{response}(a_k, a_i, (o_j,
r_j), T_{a_i,...,a_k}) \in R$ where $R$ is the set of all responses. The
values of trust along the path provide a ranking of the recommendations.
There are many ways of choosing based on such rankings; we would like to
introduce an exploratory behaviour of agents and an established way of
doing so consists in choosing randomly among all recommendations with
probabilities assigned by a logit function \citep{weisbuch00}. For this
purpose, it is convenient to first map trust into an intermediate
variable $\hat{T}$, ranging in $[-\infty,\infty]$:

\begin{eqnarray}
  \hat{T}_{a_i,...,a_k} = \frac{1}{2}\ln\left(\frac{1 + 2(T_{a_i,...,a_k}-0.5)}{1 - 2(T_{a_i,...,a_k}-0.5)}\right) \in [-\infty,\infty]
  \label{decision_making_logit-2}
\end{eqnarray}

i.e. $\hat{T}_{a_i,...,a_k} = -\infty$ for $T_{a_i,...,a_k}=0$ and
$\hat{T}_{a_i,...,a_k} = \infty$ for $T_{a_i,...,a_k}=1$. Then,

\begin{eqnarray}
  P(\textrm{response}(a_k, a_i, (o_j, r_j), T_{a_i,...,a_k})) = \frac{\exp( \beta \hat{T}_{a_i,...,a_k} )}{\sum_R \exp(\beta
    \hat{T}_{a_i,...,a_l} )} \in [0,1]
  \label{decision_making_logit-1}
\end{eqnarray}

where $\beta$ is a parameter controlling the exploratory behaviour of
agents (when $\hat{T}_{a_i,...,a_k} = \pm \infty$,
$P(\textrm{response}(a_k, a_i, (o_j, r_j), T_{a_i,...,a_k}))$ is computed
as a limit). With such transformations we achieve to have trust values
$T_{a_i,...,a_k}$ to lie in $[0, 1]$ which is required in order to
propagate them as well as negative values of $\hat{T}_{a_i,...,a_k}$ when
the trust towards an agent is very small -- otherwise, agents would keep
choosing recommendations even from untrustworthy agents with finite
probability. For $\beta=0$, the probability of choosing each response
will be the same (i.e. this is equivalent to a random choice), but for
$\beta>0$, responses with higher associated values of $T_{a_i,...,a_k}$
have higher probabilities. To decide for one of the objects, the agent
chooses randomly between all recommendations according to these
probabilities.

Now, suppose that an agent received a recommendation from another agent,
but through many paths. For example, $a_i$ may be linked to $a_k$ through
$a_j$, but also through $a_l$. Then, each of the two responses would be
assigned a probability according to eq. \ref{decision_making_logit-1}.
Since recommendations coming along paths of high trust will have a higher
probability of being chosen, this implies that recommendations coming
along paths of low trust are still part of the decision-making process,
but with much lower probability. This approach is similar to
\citep{gray03} (where only the highest path is considered, and all lower
paths are discarded) and the issue has also been discussed in
\citep{josang05}.

For benchmarking the trust-based approach of selecting recommendations,
we consider an alternative decision making strategy, namely a
\textit{frequency-based approach} without any trust relationships being
considered at all. In this approach, an agent chooses randomly among each
of the recommendations with equal probability for each of the
recommendations.

\subsection{Trust Dynamics}

In order to enable the agents to learn from their experience with other
agents, it is necessary to feedback the experience of following a
particular recommendation into the trust relationship. This is done as
follows: subsequent to an interaction, agent $a_i$ who has acted on a
rating through its neighbour, agent $a_j$, updates the value of trust to
this neighbour, based on the experience that he made. Let $o_k$ be the
chosen object. Then, assuming agent $a_i$ having profile $p_i$,
$p_i(o_k)=r_k$ is the experience that $a_i$ has made by following the
recommendation transmitted through $a_j$. It is convenient to define the
update of $T(t+1)$ in terms of an intermediate variable $\tilde{T}(t+1)$:

\begin{eqnarray}
  \tilde{T}_{a_i,a_j}(t+1) = \left\{ \begin{array}{cc}
  \gamma \tilde{T}_{a_i,a_j}(t) + (1 - \gamma) r_k & \mathrm{for}\ r_k \ge 0 \\
  (1 - \gamma) \tilde{T}_{a_i,a_j}(t) + \gamma r_k & \mathrm{for}\ r_k  < 0
  \end{array} \right.
  \label{trust_update1}
\end{eqnarray}

where $ \tilde{T}_{a_i,a_j}(0) = 0 $ and $\gamma \in [0,1]$.  Because $
\tilde{T}_{a_i,a_j} \in [-1,1]$, we have to map it back to the
interval $[0,1]$:

\begin{eqnarray}
  T_{a_i,a_j}(t+1) = \frac{1+\tilde{T}_{a_i,a_j}(t+1)}{2} \in [0,1]
  \label{trust_update2}
\end{eqnarray}

The distinction between $r_k \ge 0$ and $r_k < 0$ creates, for values of
$\gamma > 0.5$, a slow-positive and a fast-negative effect which usually
is a desired property for the dynamics of trust: trust is supposed to
build up slowly, but to be torn down quickly. The trust update is only
applied between neighbouring agents. Although the trust along pathways
between two non-neighbour agents $T_{a_i,...,a_j}$ is used for choosing a
recommendation, this is not used to establish a value of trust towards
non-neighbour agents. The trust along pathways between two non-neighbour
agents $T_{a_i,...,a_j}$ changes only as a result of changes on the links
of the path, i.e.  changes between neighbour agents.

Our intention is to keep the trust dynamics local, i.e. restrict it to
neighbours. Any other approach would require agents to maintain global
knowledge. The performance of the system results from the development of
pathways of high trust and thus is an emergent property of local
interactions between neighbouring agents.

It is important to note that -- in the current version of the model -- as
a result of the trust dynamics, trust $T_{a_i,a_j}$ evolves to a value
which reflects the similarity of agents $a_i$ and $a_j$. This is
consistent with the observation in the literature that there is a
correlation between trust and similarity \citep{golbeck06c} and that, in
a recommendation system, ``recommendations only make sense when obtained
from like-minded people exhibiting similar taste'' \citep{ziegler06}. In
fact, our mechanism could be seen as a possible explanation of this
correlation. However, as stated, there are other interpretations of trust
in different disciplines, in particular cognitive science, sociology, and
psychology.

In further extensions of the model, trust $T_{a_i,a_j}$ could include
other notions such as ``agent $a_j$ cooperated with agent $a_i$'',
``agent $a_j$ gave faithful information to agent $a_i$'', or ``agent
$a_j$ joined a coalition with agent $a_i$''. In other words,
$T_{a_i,a_j}$ could be an aggregate of different dimensions of trust,
possibly measuring the faithfulness, reliability, availability, and
quality of advice from a particular agent.

\subsection{Utility of Agents, Performance of the System}

In order to quantitatively measure the difference of the trust-based
approach of selecting recommendations as compared to the frequency-based
approach, it is necessary to define measures for the utility of agents as
well as for the performance of the system.

We define an instantaneous utility function for an agent $a_i$ following
a recommendation from agent $a_j$ on object $o_k$ at time $t$ as follows:

\vspace{-5mm}

\begin{eqnarray}
  u(a_i, t)=r_i 
  \label{utility_of_agents_concrete_instantaneous}
\end{eqnarray}

where agent $a_i$'s profile determines $p_i(o_k)=r_i$. We consider the
performance of the system to be the average of the utilities of the
agents in the system:

\vspace{-5mm}

\begin{eqnarray}
  \Phi(t)=\frac{1}{N_A}\sum_{a_i \in S_A}{u(a_i, t)}
  \label{performance_of_the_system_instantaneous}
\end{eqnarray}

This gives us a measure for quantitatively comparing the difference that
the trust-based approach makes towards the frequency-based approach, both
on the micro-level of an agent and the macro-level of the system.

\section{Results}

One of the most important results of the model is that the system
self-organises in a state with performance near to the optimum.  Despite
the fact that agents only consider their own utility function and that
they do not try to coordinate, long paths of high trust develop in the
network, allowing agents to rely on recommendations from agents with
similar preferences, even when these are far away in the network.
Therefore, the good performance of the system is an emergent property,
achieved without explicit coordination.

Three quantities are particularly important for the performance of the
system: the network density, the preference heterogeneity among the
agents, and the sparseness of knowledge. The core result is that
recommendation systems in trust-based networks outperform frequency-based
recommendation systems within a wide range of these three quantities:

\begin{itemize}
\item \textit{Network density}: if the network is very sparse, agents
  receive useful recommendations on only a fraction of the items that
  they send queries about; the denser the network, the better the
  performance, but above a critical threshold for the density, the
  performance stabilises. The proximity of this value to the optimum
  depends on the other two quantities.
\item \textit{Preference heterogeneity}: if the preferences of agents are
  homogeneous, there is no advantage for filtering the recommendations;
  however, if the preferences of agents are all different, agents cannot
  find other agents to act as suitable filters for them. In between, when
  preferences are heterogeneous, but `not too much', the system
  performance can be near to the optimum.
\item \textit{Knowledge sparseness}: when knowledge is dense ($N_c$
  and/or $N_p$ small), it is easy for an agent to receive recommendations
  from agents with similar preferences. In the extreme situation in
  which, for each category there is only one expert with any given
  preference profile, agents can receive useful recommendations on all
  categories only if there exists a high-trust path connecting any two
  agents with the same profile. This is, of course, related to the
  density of links in the network.
\end{itemize}

The performance of the system thus depends, non-linearly, on a
combination of these three key quantities. Under certain assumptions, the
model can be investigated analytically and in a mean-field approach it is
possible to make quantitative predictions on how these factors impact the
performance. These results are presented in subsection
\ref{sec:analytical_appoximation}. In subsection
\ref{sec:computer_simulations}, we illustrate the properties of our
recommendation system by describing the results of multi-agent
simulations of the model. As a benchmark, we compare the trust-based
recommendation system to a frequency-based recommendation system.

\subsection{Analytical Approximation}
\label{sec:analytical_appoximation}

In the following, we derive an expression for the performance of the
system as a function of the frequency and heterogeneity of profiles
across agents. We proceed as follows. We first introduce the notion of
similarity, $\omega$ of profiles. We then show, in the limit of a
mean-field approximation, that the fix points of trust between two agents
are a function of the similarity of their profiles. We then derive the
value of the critical threshold for the network density above which a
subset of agents with the same profile is expected to form a connected
component.  Above this threshold, agents with the same profile can
receive recommendations on all categories covered by the expertise of
such a subset of agents. Under this hypothesis, and in the stationary
regime for the trust dynamics, the expected utility of an agent can
easily be computed, again in a mean-field approximation, based on the
decision making dynamics used to choose among recommendations.

As common in the literature, the similarity between two profiles
$p_i,p_j$ is defined as
\begin{eqnarray} 
  \omega_{i,j}=\frac{1}{N_O}\sum_{o_k\in S_O} 1-\abs{p_i(o_k)-p_j(o_k)} 
  \label{eq:avg overlap}
\end{eqnarray}

The similarity of two agents is, for instance, $1$ if their preferences
over the products are identical, and $-1$ if their preferences over the
products are always opposite, and $0$ if half of their preferences are
identical and half are opposite.

Suppose there are only two profiles $p_1,p_2$ in the population. If
profiles are evenly distributed among agents ($n_1=1/2$), then the
expected value of $\omega$ over a large set of pairs of profiles is
$\mean{\omega}=0$. If instead, agents have only one profile, $p_1=1$,
then trivially $\mean{\omega}=1$.

\subsubsection{Trust Dynamics}

Consider the dynamics for the update of trust of agent $a_i$ towards a
neighbouring agent $a_j$ (eq. \ref{trust_update1}). Assume the two agents
have profiles $p_m$ and $p_n$, respectively (since the number of agents
and of profiles are different, we don't use $p_i$ and $p_j$ for the
profiles as this would suggest that there exactly as many profiles as
agents). Let us then focus first only on recommendations coming directly
from $a_j$.  Eq. \ref{trust_update1} is a stochastic equation because
recommendations are provided by $a_j$ to $a_i$ on objects of randomly
chosen categories. In a mean-field approximation, we replace the
stochastic term, $r_k$, with its average over time, which, by definition,
is $\omega_{m,n}$.  It is then straightforward to check that the fix
point of both cases of eq.  \ref{trust_update1} is $\omega_{m,n}$. By
the time agent $a_i$ has developed a value of trust towards $a_j$ close
to $\omega_{m,n}$, $a_j$ has done the same with its own neighbours.
In particular, if $\omega_{m,n}$ is close to 1, then only
recommendations from neighbours $a_w$ of $a_j$ towards which $a_j$ has
developed high trust, are associated with high values of trust along the
path $a_w,a_j,a_i$.  The same holds, by induction, for longer paths.
Therefore, we can extend the mean-field approximation also to the general
case of recommendations received by $a_i$ indirectly through $a_j$.

\subsubsection{Random Graph Structure and Critical Density}
\label{subsec:analyt_graphs}

It is known that in a random graph of $N$ nodes and $\ell$ links, a
\textit{giant connected component} appears for values of $\ell>(N-1)/2$,
meaning that the probability that the network is connected tends to $1$
for large $N$ (and correspondingly large $\ell$)
\citep{erdos59,bollobas85}. Equivalently, above this threshold, there is
at least one path between any two randomly chosen nodes. In our model,
agents are connected in a random graph and have different preference
profiles, distributed randomly according to some frequency distribution.
We can then ask what is the critical density of links (randomly drawn
among agents of any profile) in the network such that there is (in the
limit of many agents) at least one path between any two agents with the
same profile. In this situation, a querying agent is able to receive
recommendations from all other agents of the same profile along paths
which involve only agents of the same profile. If $n_i$ is the frequency
of agents of profile $p_i$, we denote $\ell_{i,i}=\ell n_i^2$ to be the
number of links among any two agents with same profile $p_i$. The
condition for the existence of a giant component of agents with profile
$p_i$ is $\ell_{i,i} > \frac{N-1}{2}$, which implies $\ell >
\frac{N-1}{2n_i^2}$. For instance, for two profiles with frequency
$n=0.5$, this formula leads to $\ell = 2 (N-1)$. In other words, the
smaller the frequency of profile $p_i$, the higher the critical number of
links $\ell$ above which agents with profile $p_i$ become connected in a
giant component.

\subsubsection{Performance}
\label{subsec:analyt_performance}

As described in the decision making process, at each time step, as a
result of a query for a given category, an agent $a_i$ receives a set of
ratings associated with values of the trust along the paths from which
the responses come from. Each rating is selected with a probability given
by eq. \ref{decision_making_logit-1}. Over time, the agent sends many
queries. Let $R$ be the set of all responses $k$ it receives over time.
The expected value of the rating $r$, hence of the utility $u$ of the
agent, is then:
\begin{eqnarray} 
   E(u)=E(r)=\sum_{k \in R} r_k P_{k}=
    \frac{\sum_{k \in R} r_k \exp(\beta \hat{T}_{k})}{\sum_{k \in R}
      \exp(\beta \hat{T}_{k})}
    \label{eq:phi_1}
\end{eqnarray}

The set R contains responses sent by many agents with different profiles.
We can group the set R by the set $S_p$ of profiles of such agents. In a
mean-field approach we can then replace the ratings experienced by the
agent $a_i$, with profile $p_q$, through following recommendations of
other agents with profile $p_s$, with its average value $\omega_{q,s}$.
In the same spirit, we can also replace the value of trust towards all
agents with profile $p_s$ by $\omega_{q,s}$. Then, eq.
\ref{decision_making_logit-2} implies that $\exp(\beta \hat{T})$ is
approximated with the value
$(\frac{1+\omega}{1-\omega})^\frac{\beta}{2}$. This approximation is
well-justified for first neighbours. For the other agents, it is less
accurate, but it may be expected to hold if the network is well above the
density threshold and in the stationary regime in which trust paths have
already developed. The expected utility of an agent, which, in the long
run, coincides with the expected value of the performance of the system,
is then:

\begin{eqnarray} \label{eq:phi_final}
   E(\Phi)=
    \frac{\sum_{s \in S_p}\omega_{q,s}(\frac{1+\omega_{q,s}}{1-\omega_{q,s}})^\frac{\beta}{2}}
    {\sum_{s \in S_p}(\frac{1+\omega_{q,s}}{1-\omega_{q,s}})^\frac{\beta}{2}}=
 \frac{\sum_{\omega} \omega (\frac{1+\omega}{1-\omega})^\frac{\beta}{2}\nu(\omega)}
 {\sum_{\omega} (\frac{1+\omega}{1-\omega})^\frac{\beta}{2}\nu(\omega)}
\end{eqnarray}

where the second expression is obtained as follows: we group the set
$S_p$ by the values of similarity between the profile of the querying
agent and the profiles of the recommending agents. Because in a pair of
querying-responding agents there is a finite number of combinations of
profiles, and their probability of occurrence depends the relative
frequency of each profile in the population (profiles are assigned
randomly to the agents). Therefore, the probability of occurrence of each
value of similarity $\omega$, $\nu(\omega)$ is known by construction.
Each term $(\frac{1+\omega}{1-\omega})^\frac{\beta}{2}\nu(\omega)$
represents the probability of an agent choosing the recommendation from
an other agent with a given similarity value $\omega$, multiplied by the
probability that such a similarity value occurs among the querying agent
and the recommenders. This formula allows to predict the expected utility
of the system as a function of the distribution of the profiles of
preferences among the agents. The formula holds in the regime in which
each subset of agents of the same profile form a connected component and
their joint expertise covers all the categories. For instance, if we
consider two profiles in the system $p_1$ and $p_2$, with frequency $n_1$
and $1-n_1$, the probability that a pair of agents consists of both
$p_1$, or both $p_2$, or mixed is, respectively: $(n_1)^2$,
$(n_2)^2=(1-n_1)^2$ and $2(n_1)(1-n_1)$.  The corresponding values of
$\omega$ are ${1,1,-1}$.

In the absence of trust (i.e., $\beta=0$), eq. \ref{eq:phi_final} reduces
to the expression of the expected value of $\omega$, yielding
$\Phi=4n_1^2-4n_1 +1$. In presence of trust (i.e., $\beta>0$), the term
with $\omega$ close to $1$ dominates, thus yielding $\Phi \approx 1$.
These results will be confirmed empirically in the next subsection.

\subsection{Computer Simulations}
\label{sec:computer_simulations}

For the simulations we have used the following parameters to the model:
we consider $N_a=100$ agents, and the simulations are averaged over
$N_r=100$ runs.  The size of each category is the same and we vary $N_c
\in \{10,...,50\}$ and $N_p \in \{2,4,6\}$; $N_o$ is usually adjusted
such that there are at least $2$ objects in each category.  Profiles are
distributed such that the sum over a profile is $0$ on average -- across
the profile, categories, and agents.  Each agent is an expert on $1$
category. Further, for the social network we assume a random directed
graph with a given number of agents, $N_a$, and a given total number of
links, $\ell$. The \textit{network density} is then defined as
$d=\ell/N_a(N_a-1)$. Agents are connected randomly with respect to their
profile.

\begin{figure}[htbp]
  \centering
  \begin{minipage}[t]{0.4\textwidth}
    \includegraphics[width=\textwidth]{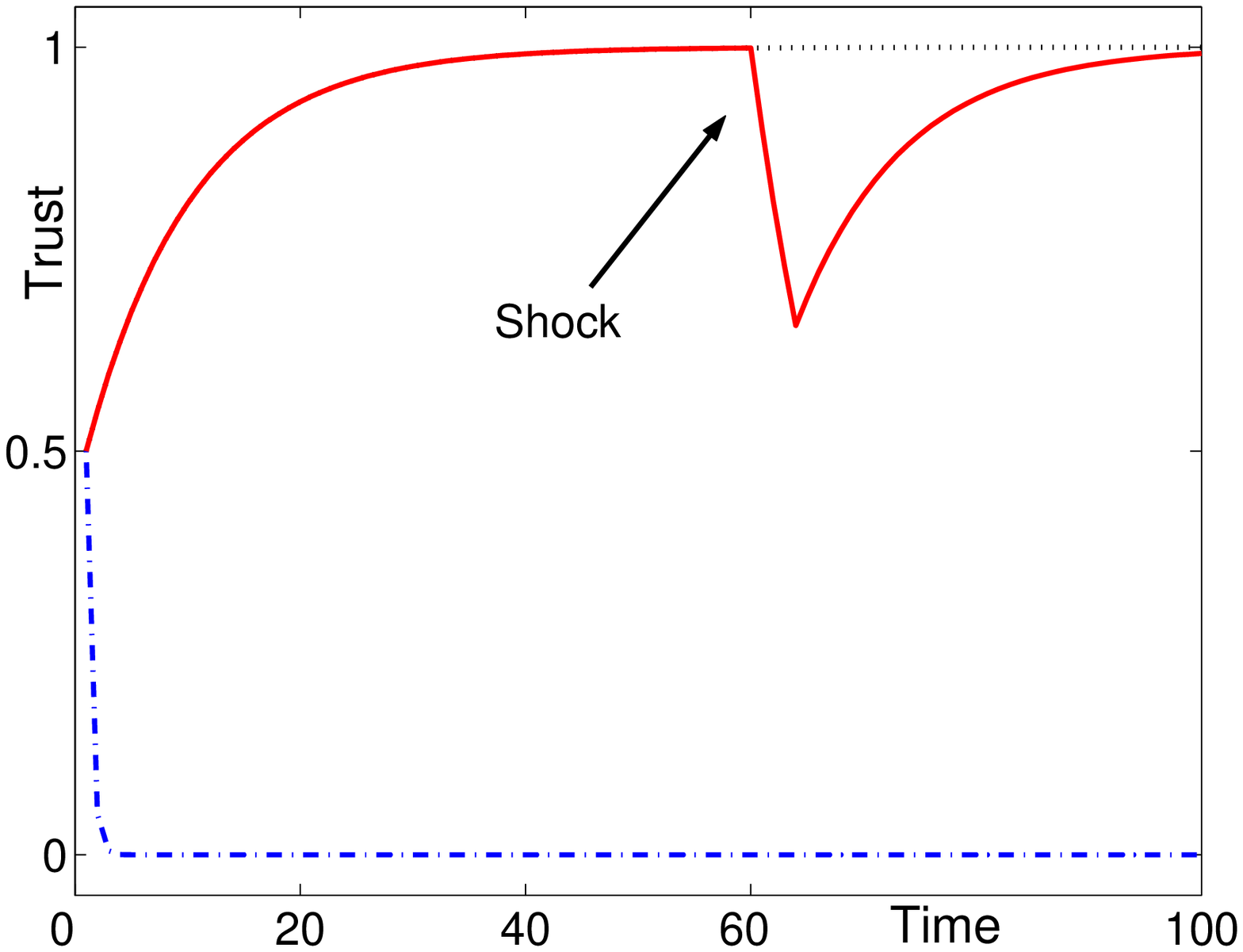}
  \end{minipage}
  \begin{minipage}[t]{0.4\textwidth}
    \includegraphics[width=\textwidth]{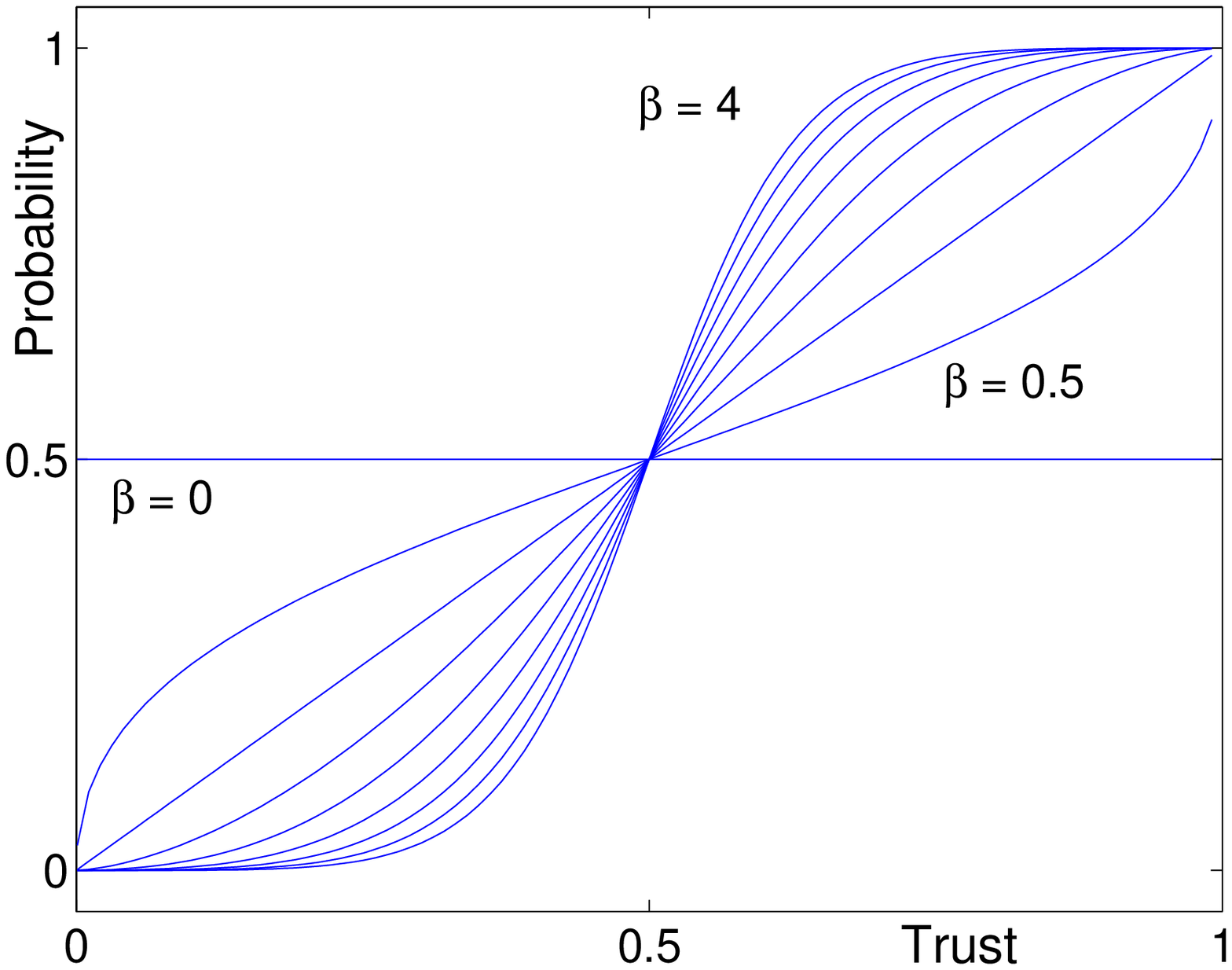}
  \end{minipage}
  \caption{Dynamics of Trust and Logit Function. Left: slow-positive
    fast-negative dynamics of trust. Trust between two agents of the same
    profile (black dotted line), between two agents of opposite profiles
    (blue dashed line). In case that an agent recommends an object that
    is rated negatively, trust drops quickly and recovers slowly (red
    solid line). Right: impact of the choice of the exploration parameter
    $\beta$ on the decision making. The slope of the sigmoid-like
    function increases for increasing values of $\beta$.}
  \label{fig:trust_evol_and_proba}
\end{figure}

Figure \ref{fig:trust_evol_and_proba} (left) shows that the update rule
of trust as described by eq. \ref{trust_update1} and \ref{trust_update2}
produces the desired slow-positive fast-negative dynamics. Trust between
two agents of the same profile evolves to $1$ (black dotted line,
partially covered by the red solid line). Trust between two agents of
opposite profiles evolves to $0$ (blue dashed line). In case that an
agent recommends an object that is rated negatively, trust drops quickly
and recovers slowly (red solid line). The probability of choosing a
recommendation depends critically on the parameter $\beta$, which
controls the exploratory behaviour of agents, as shown in Figure
\ref{fig:trust_evol_and_proba} (right).

\begin{figure}[htbp]
  \centering
  \begin{minipage}[t]{0.4\textwidth}
    \includegraphics[width=\textwidth]{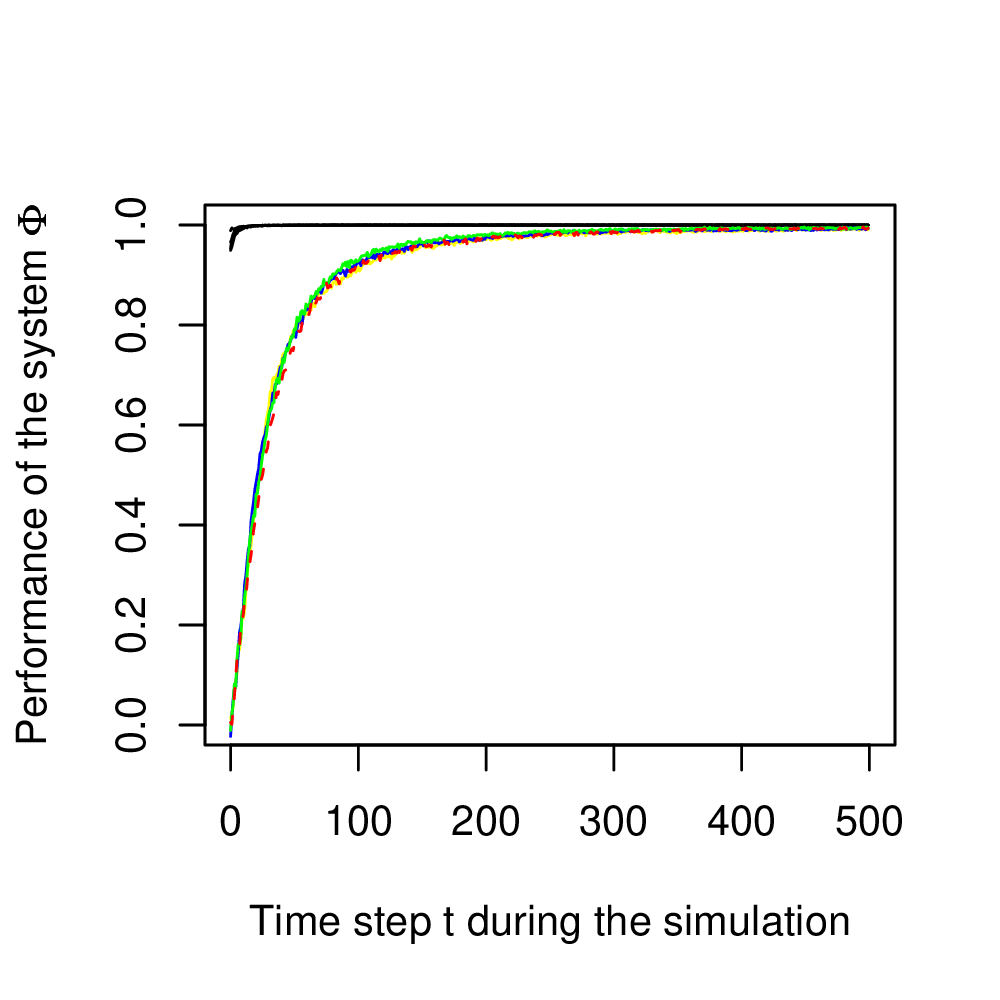}
  \end{minipage}
  \begin{minipage}[t]{0.4\textwidth}
    \includegraphics[width=\textwidth]{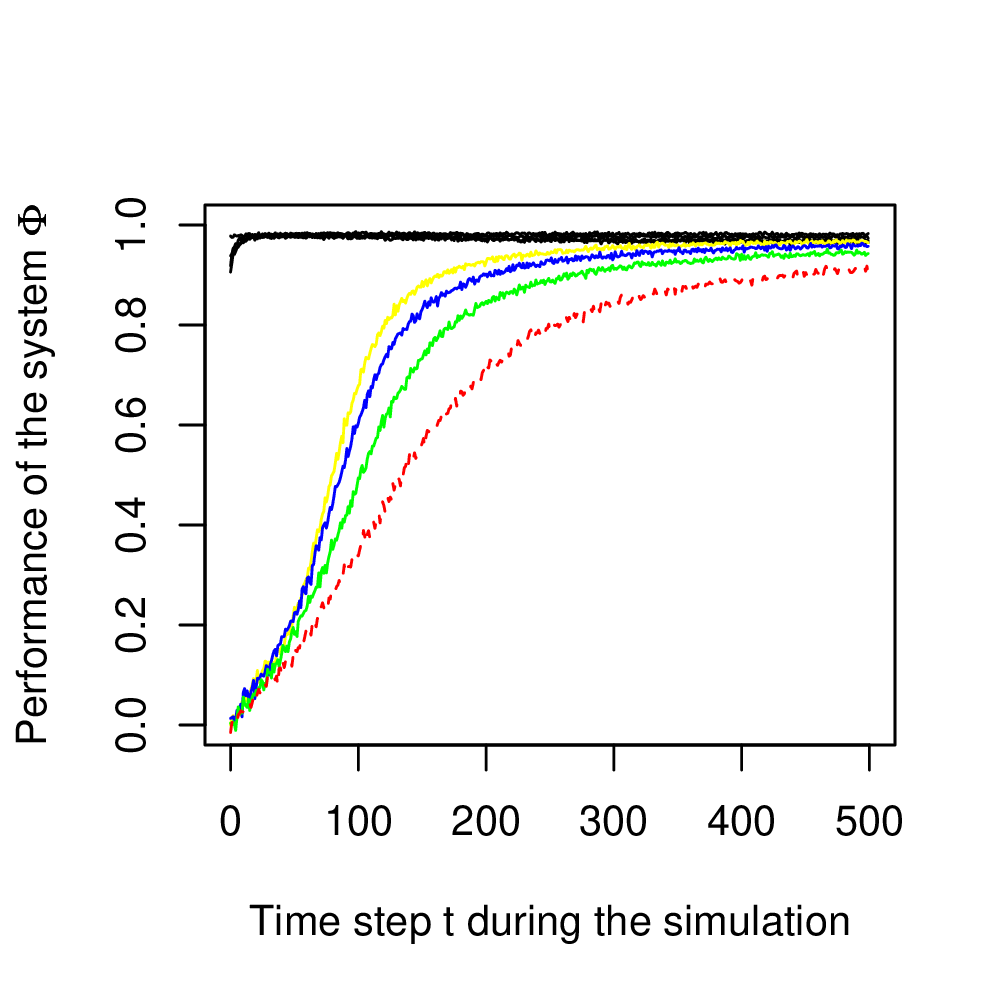}
  \end{minipage}
  \caption{Performance vs. Time for $N_c=10$ (left) and $N_c=50$ (right).
    Over time, performance approaches the optimum -- with learning (top
    black line), this process is accelerated. Different curves correspond
    to different values of $\gamma \in \{0.2, 0.4, 0.6, 0.8\}$.
    Increasing values of $\gamma$ lead to curves approaching the optimum
    faster (corresponding colours: red, green, blue, yellow).}
  \label{fig:plot-1and2-no-learning-is-required}
\end{figure}

Over time, each agent develops a value of trust towards its neighbours
which reflects the similarity of their respective profiles.  After some
time, paths of high trust develop, connecting agents with similar
profiles. As a result, the performance of the system, as defined in eq.
\ref{performance_of_the_system_instantaneous} increases over time and
reaches a stationary value which can approach the optimum, as shown in
Figure \ref{fig:plot-1and2-no-learning-is-required}, where the curves
correspond to different values of $\gamma$. Increasing values of $\gamma$
lead to curves approaching the optimum faster.

We have also simulated a situation in which, prior to the start of the
dynamics, there is a learning phase in which the agents explore only the
recommendations of their direct neighbours on the categories that these
claim to be expert on. This way, the trust dynamics already start from a
value deviating from the neutral point of $0.5$ and closer to one of the
fix points (see eq. \ref{trust_update1}). In this case, the performance
is optimal from the beginning on (top black curve).  Interestingly, the
system evolves, even in the normal dynamics, to the same value that is
reached with the learning phase, supporting the idea that the optimal
performance is an emergent behaviour of the system.

\begin{figure}[htbp]
  \centering
  \begin{minipage}[t]{0.4\textwidth}
    \includegraphics[width=\textwidth]{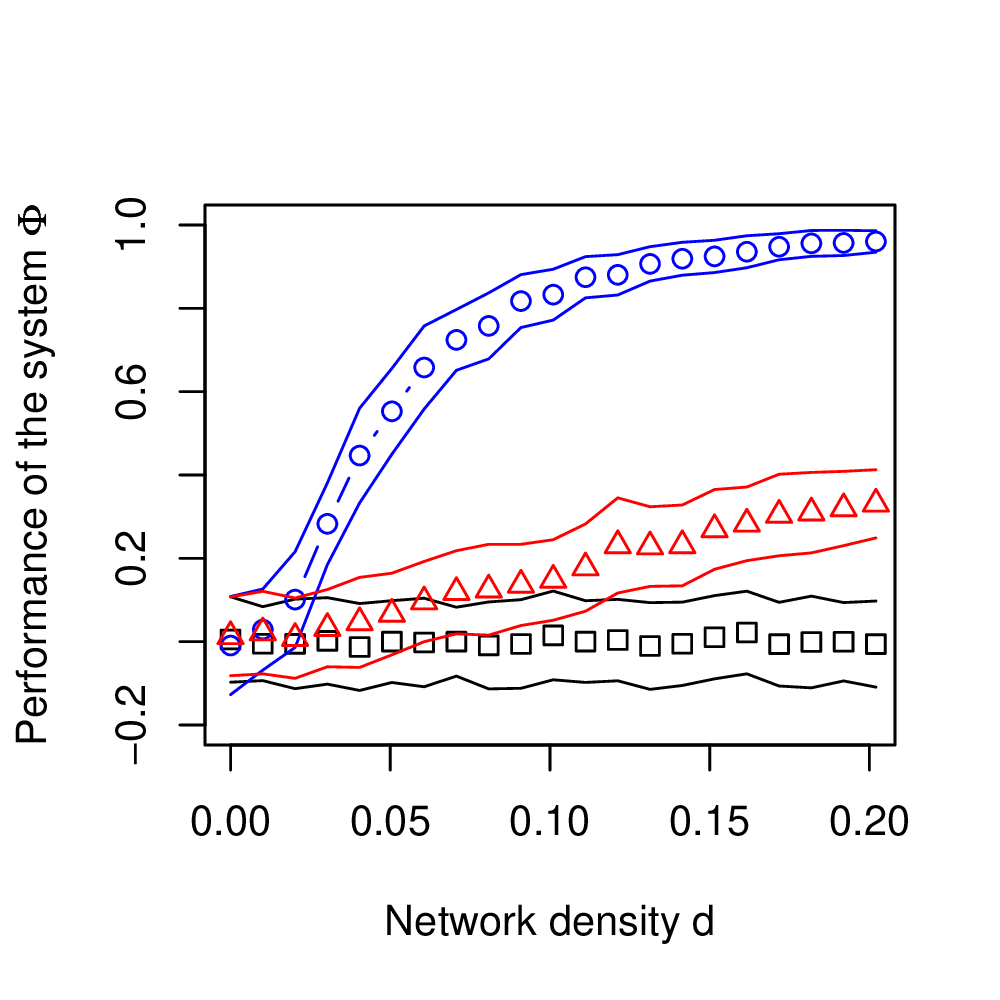}
  \end{minipage}
  \begin{minipage}[t]{0.4\textwidth}
    \includegraphics[width=\textwidth]{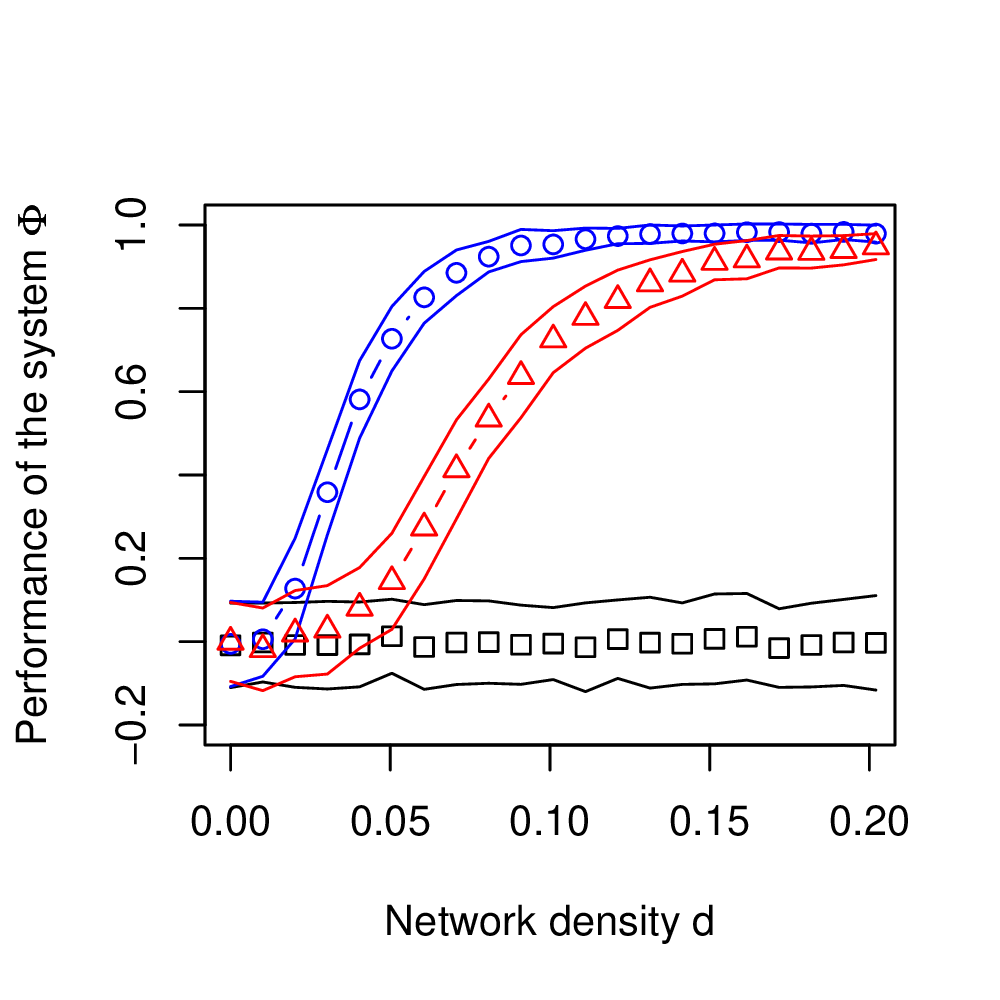}
  \end{minipage}
  \caption[]{Performance vs. density for different $N_c$ (in blue circles
    and red triangles, $10$ and $50$ categories, respectively).
    Incomplete search (left) and complete search (right). For sparse
    knowledge, the complete search performs much better than the
    incomplete search. The benchmark of a frequency-based system is
    represented by black squares.}
  \label{fig:plot-3and4-exhaustive-search-is-required}
\end{figure}

In the model description, we have described two types of search. Figure
\ref{fig:plot-3and4-exhaustive-search-is-required} -- the performance
$\Phi$ of the system plotted against increasing values of density $d$ in
the network -- shows that the search type becomes important when the
knowledge is sparse. We notice a sigmoid shape which would become steeper
for systems with larger numbers of agents. We consider different $N_c$,
corresponding to levels of sparseness of knowledge (in blue circles and
red triangles, $10$ and $50$ categories, respectively, $N_p=2$).  With
the incomplete search algorithm, the performance deteriorates. With the
complete search algorithm, the system reaches the optimal performance
even in the case of maximally sparse knowledge ($50$ categories means
that there is only $1$ expert from each profile in each category). In
both plots of Figure \ref{fig:plot-3and4-exhaustive-search-is-required},
the black squares correspond to the frequency-based recommendation
system used as benchmark. In fact, without trust, the performance is $0$
on average, because random choices lead to an equal distribution of
``good'' and ``bad'' objects (with respect to profiles).

\begin{figure}[htbp]
  \centering
  \begin{minipage}[t]{0.4\textwidth}
    \includegraphics[width=\textwidth]{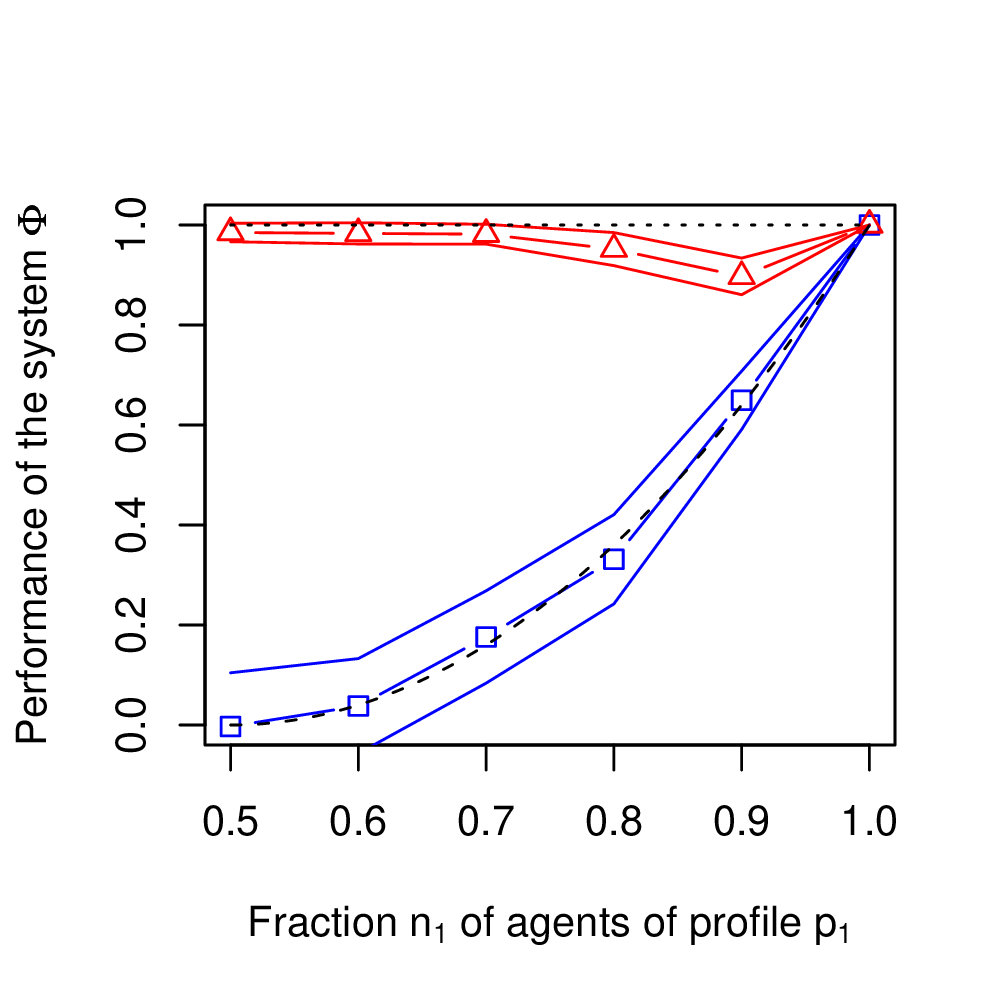}
  \end{minipage}
  \begin{minipage}[t]{0.4\textwidth}
    \includegraphics[width=\textwidth]{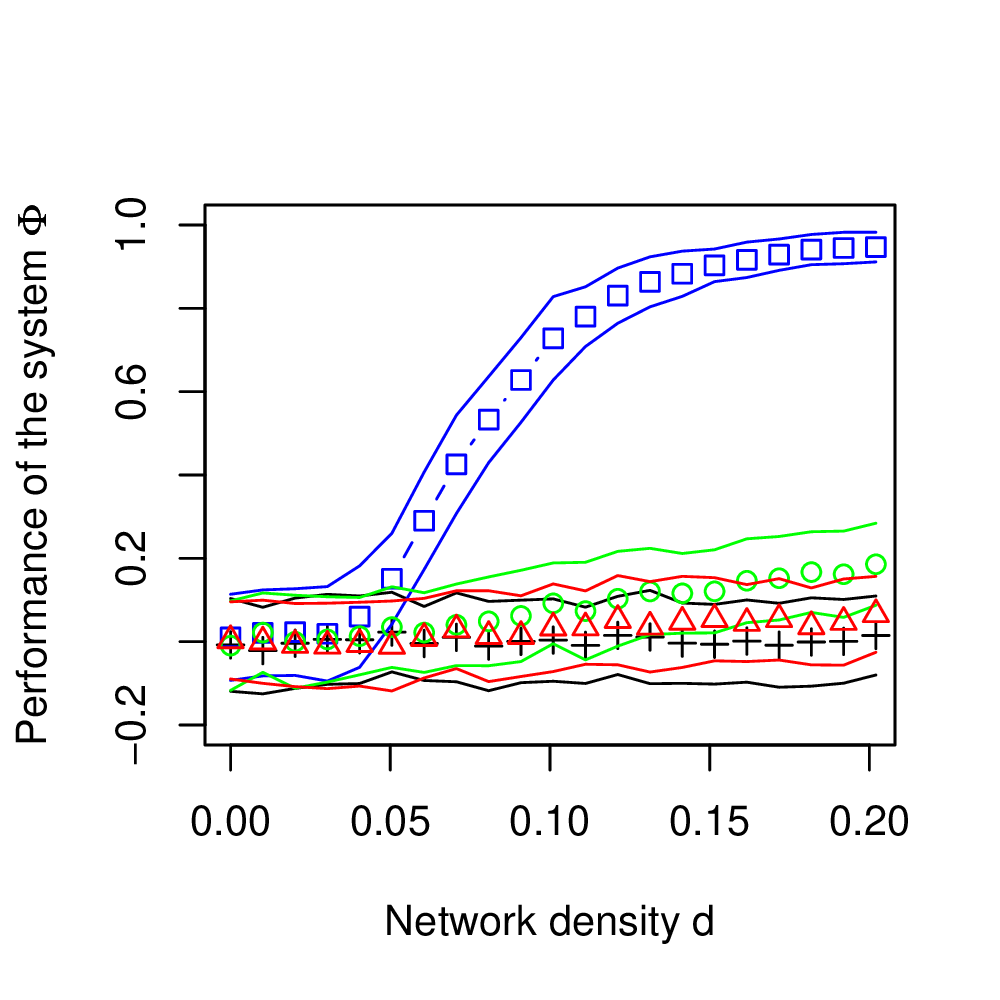}
  \end{minipage}
  \caption{Effect of heterogeneity on performance. The trust-based
    approach performs well also in very homogeneous systems; in the
    extreme case of very heterogeneous systems, performance drops. Left:
    Performance as a function of the fraction $n_1$ of agents with
    profile $p_1$, with trust (red triangles) and without trust (blue
    squares). Right: Performance as a function of network density $d$
    with different numbers of profiles $N_p$ (blue squares: $N_p=2$,
    green circles: $N_p=4$, red triangles: $N_p=6$). The benchmark of a
    frequency-based system is represented by black crosses.}
  \label{fig:plot-5and6-trust-and-heterogeneity}
\end{figure}

We now illustrate the role of preference heterogeneity. We consider first
the case in which there are two possible, opposite, profiles in the
population, say $p_1$ and $p_2$.  We define the fraction of agents
characterised by the first profile as $n_1$.  In Figure
\ref{fig:plot-5and6-trust-and-heterogeneity} (left), we plot the
performance of the system with and without trust (red triangles and blue
squares, respectively) against increasing values of $n_1$. When $n_1=0.5$
there is an equal frequency of both profiles, while when $n_1=1$ all
agents have the first profile. For the system without trust, the
performance increases for increasing $n_1$. In fact, despite that choices
are random, agents receive recommendations which are more and more likely
to match the preferences of the majority. On the other hand, the minority
of agents with the profile $p_2$ are more and more likely to choose wrong
recommendations, but their contribution to the performance of the system
decreases. The simulation results are in good agreement with the
predictions obtained in the analytical approximation (black dotted
lines), eq.  \ref{eq:phi_final}. For the system with trust the
performance is almost unchanged by the frequency. This very strong result
has the following explanation: The social network is a random graph in
which agents have randomly assigned profiles. Agents assigned to $p_2$
decrease in number, but, as long as the minority, as a whole, remains
connected (there is a path connecting any two such agents) they are able
to filter the correct recommendations. At some point the further
assignment of an agent to $p_1$ causes the minority to become
disconnected and to make worse choices. In the simulations, this happens
when $n_1=0.9$ and $n_2=0.1$.  Another way of investigating the role of
heterogeneity of preferences is to consider an increasing number of
profiles in the population, each with the same frequency. In the extreme
case in which, for each category there is only one expert with any given
preference profile, the performance, at constant values of network
density $d$, drops dramatically, as shown in Figure
\ref{fig:plot-5and6-trust-and-heterogeneity} (right).

\section{Extensions}

So far, we have made the assumptions that (1) agents are self-interested
in the sense of bounded rationality, but do not act randomly, selfishly,
or maliciously and that (2) the social network of agents is fixed and
does not change over time -- no agents join or leave the networks and no
links are rewired, added, or dropped. In reality, both of these
assumptions need to be relaxed, so in further work, we plan to
investigate the behaviour of the system with respect to these issues.

\subsection{Evolving Social Network}

Considering a fixed network between agents does not appropriately depict
reality; usually, \textit{social networks evolve over time} with links
being created and deleted at each time step. For example, the network
could evolve in the following manner: at certain intervals over time,
each agent $a_i$ randomly picks one of its links -- e.g., to agent $a_j$
-- and rewires it to a random other agent in the network or keeps it,
both with a certain probability. Of course, it would make sense to tie
this probability to the level of trust the agent has on the particular
link considered for rewiring:

\vspace{-3mm}
\begin{eqnarray}
  P(\mathrm{rewire}) = 1 - T_{a_i,a_j} \\
  P(\mathrm{keep})   = T_{a_i,a_j}
  \label{rewiring_of_links}
\end{eqnarray}

i.e. $P(\mathrm{rewire}) + P(\mathrm{keep})=1$.

\begin{figure}[htbp]
  \centering
  \begin{minipage}[t]{0.3\textwidth}
    \centering (a) $t=t_{\mathrm{start}}$, $\beta=0$ 
  \end{minipage}
  \begin{minipage}[t]{0.3\textwidth}
    \centering (b) $t=t_{\mathrm{end}}$, $\beta=0$
  \end{minipage}\\
  \begin{minipage}[t]{0.3\textwidth}
    \includegraphics[width=\textwidth]{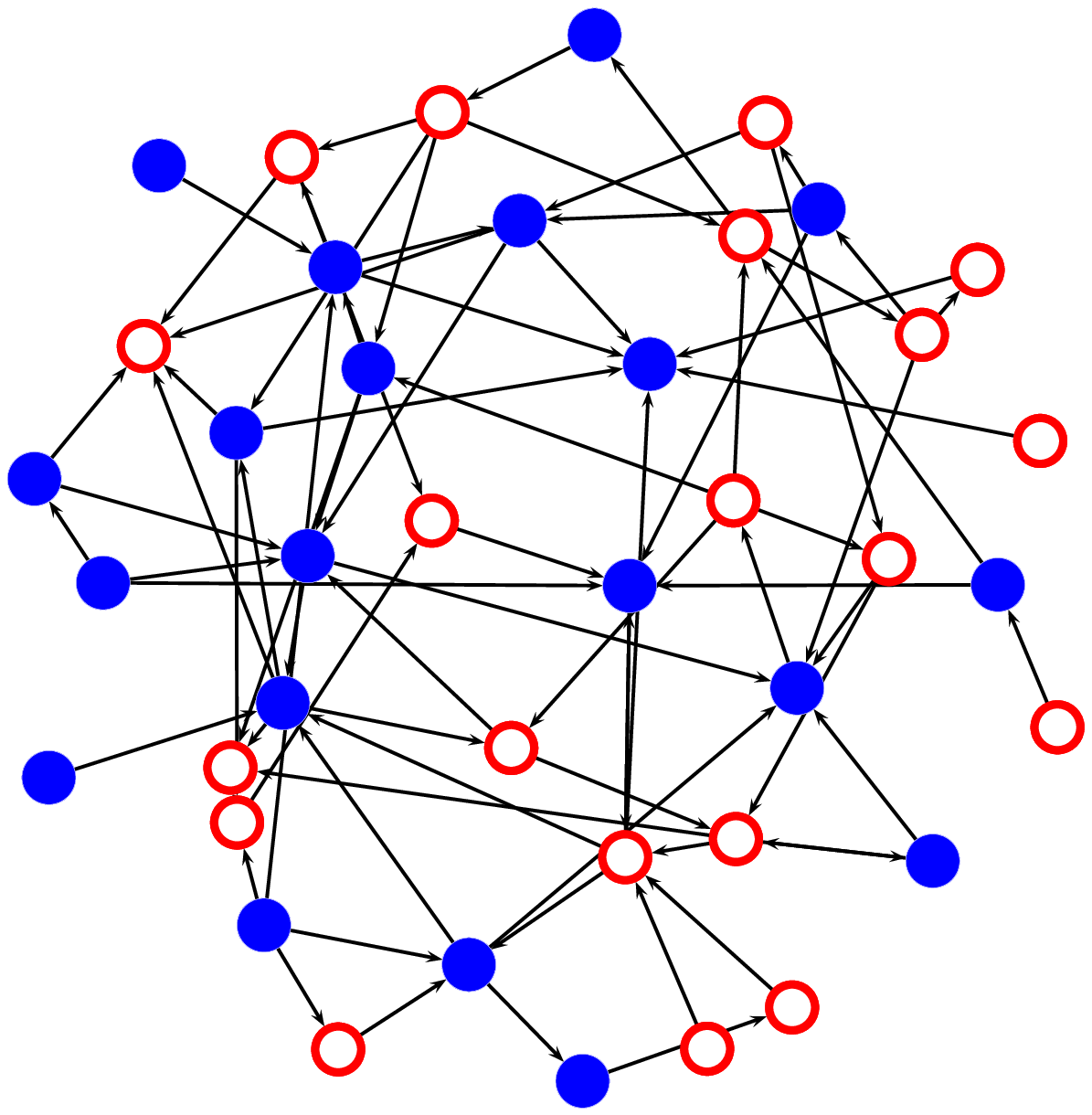}
  \end{minipage}
  \begin{minipage}[t]{0.3\textwidth}
    \includegraphics[width=\textwidth]{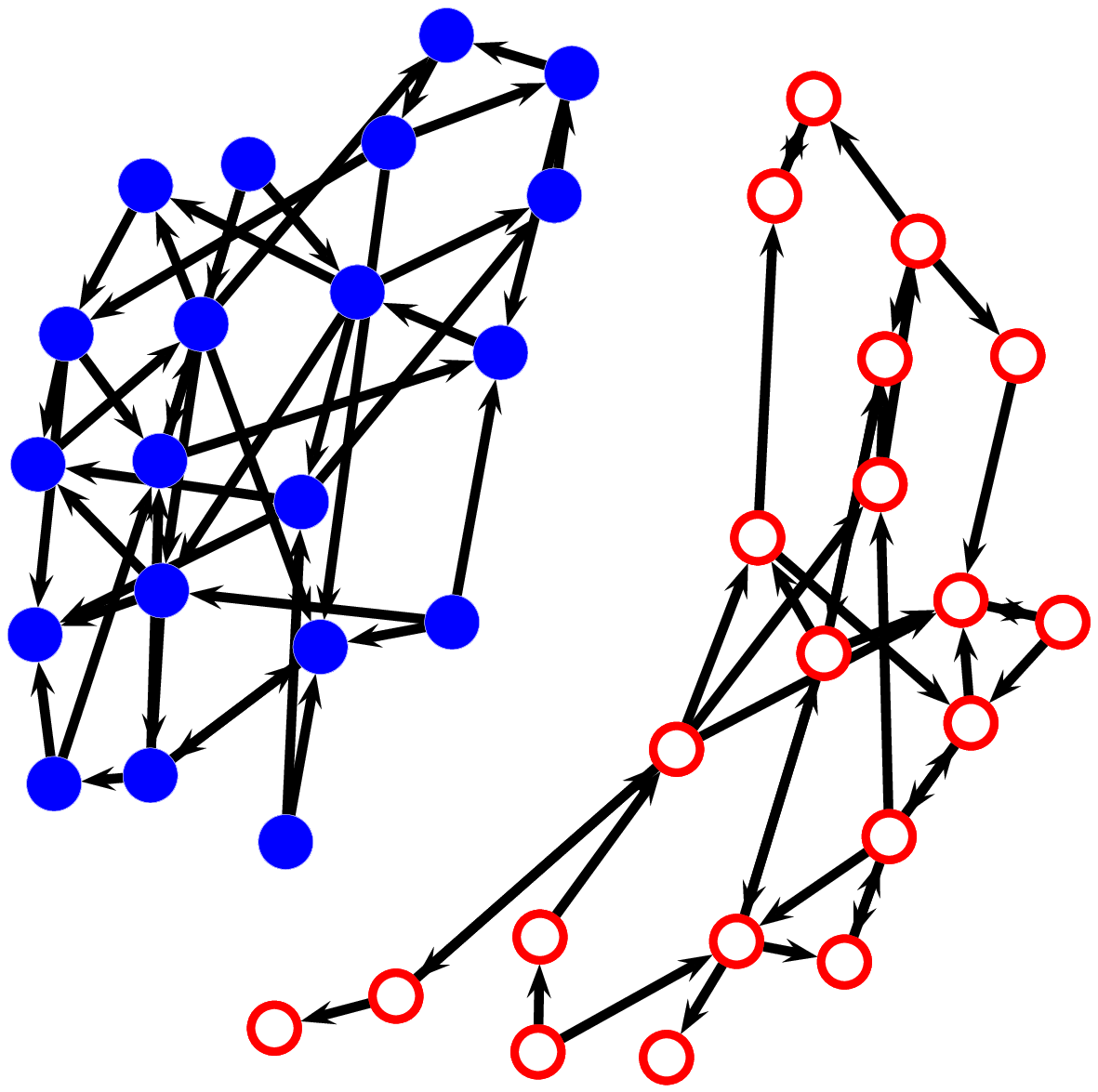}
  \end{minipage}\\
  \begin{minipage}[t]{0.3\textwidth}
    \centering (c) $t=t_{\mathrm{start}}$, $\beta=1$
  \end{minipage}
  \begin{minipage}[t]{0.3\textwidth}
    \centering (d) $t=t_{\mathrm{end}}$, $\beta=1$
  \end{minipage}\\
  \begin{minipage}[t]{0.3\textwidth}
    \includegraphics[width=\textwidth]{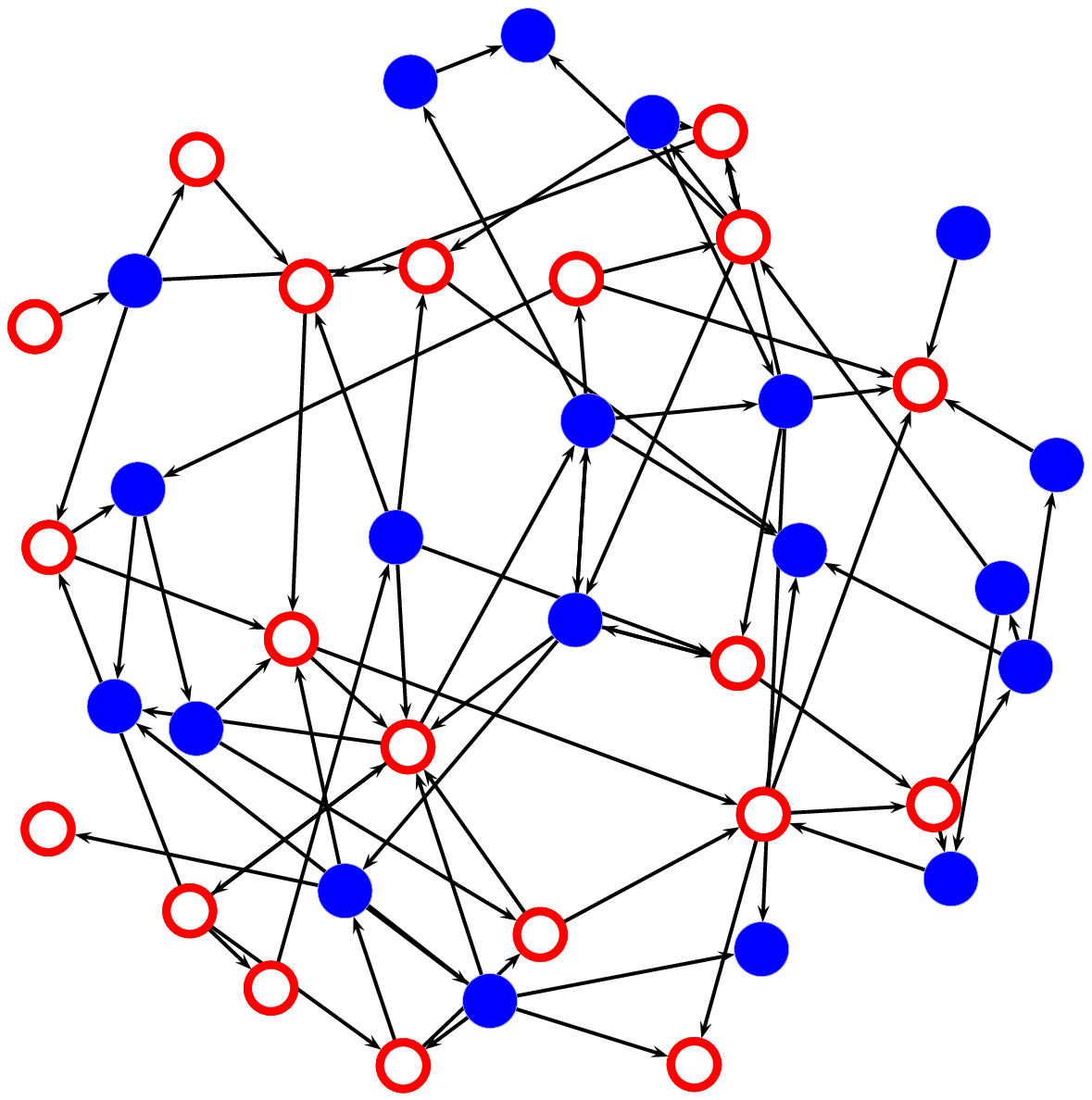}
  \end{minipage}
  \begin{minipage}[t]{0.3\textwidth}
    \includegraphics[width=\textwidth]{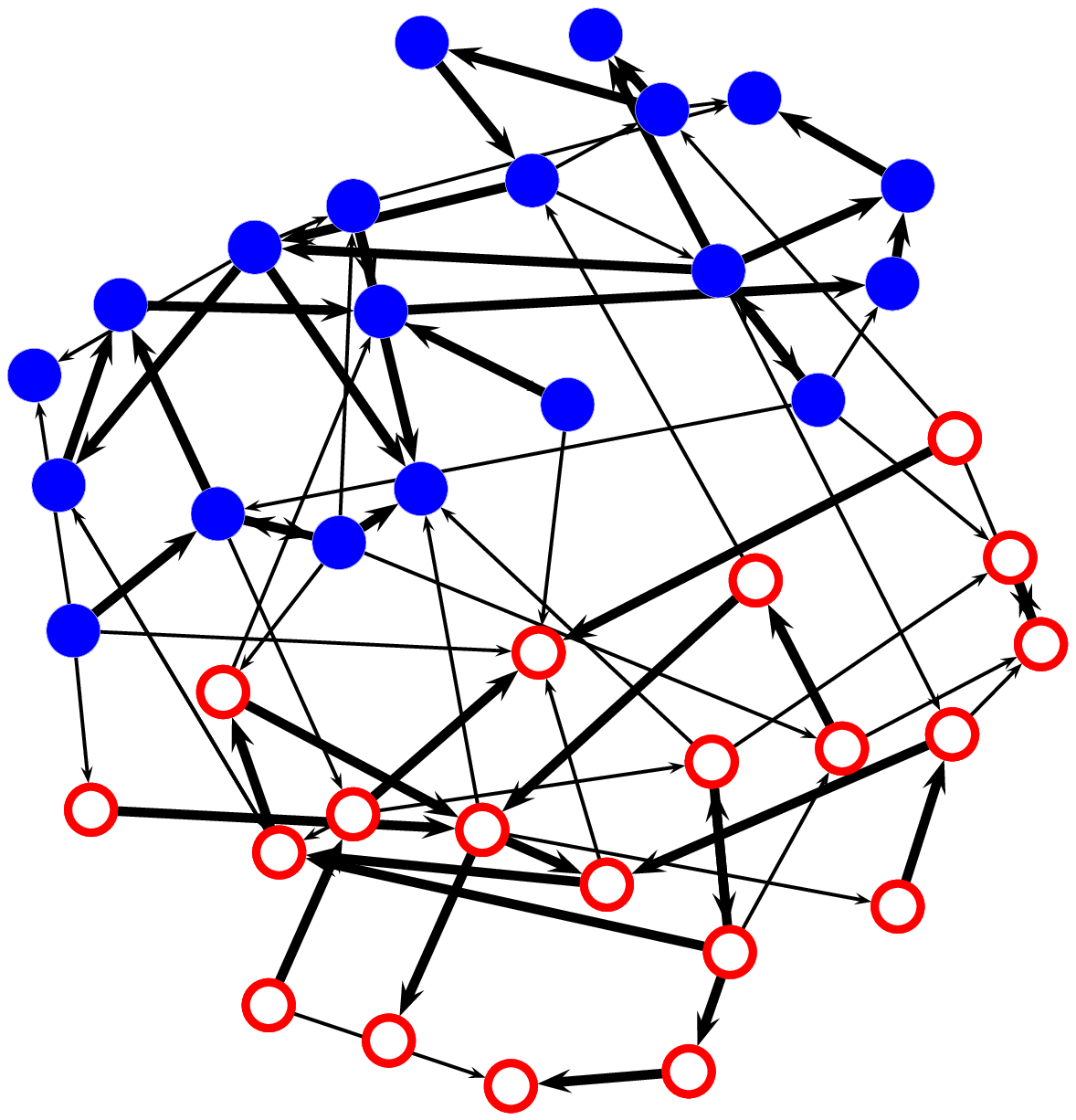}
  \end{minipage}
  \caption[Evolving Social Network]{Snapshots of the evolution of a
    network of 40 agents in 2 profiles and 80 links at time
    $t=t_{\mathrm{start}}$ and $t=t_{\mathrm{end}}$ for $\beta=0$ and
    $\beta=1$, respectively. When $\beta=0$, disconnected clusters of
    agents with the same profile form, when $\beta=1$, interconnected
    clusters of agents with the same profiles form. For $\beta>0$, agents
    develop stronger ties to agents of the same profile than to agents of
    different profiles.}
  \label{fig:disconnected-clusters-initial}
\end{figure}

Figure \ref{fig:disconnected-clusters-initial} shows how snapshots of the
evolution of a sample network of agents at different stages for different
values of $\beta$ look when applying this mechanism. Note the random
graph structure at $t=t_{\mathrm{start}}$ and the community fragmentation
at $t=t_{\mathrm{end}}$. This illustrates the dilemma between exploration
and exploitation faced by the agents. For $\beta=0$, agents choose
randomly, thus performing worse, but they explore many of the other agents
repetitively and their trust relationships converge to the steady state
of the trust dynamics of Equations \ref{trust_update1} and
\ref{trust_update2}. Then, over time, links with low trust are rewired
and links with high trust are kept. This leads to the emergence of two
disconnected clusters. Eventually, subsequent to the formation of
clusters, such agents will perform well, as any recommendation will come
from an agent of the same profile. For $\beta=1$, agents choose according
to the strength of trust relationships, thus performing better, and they
are able to exploit their knowledge. However, they exploit stronger links
while not even exploring weaker ones. This results in clustering, but
with interconnections between clusters. As networks in reality are
evolving, it is important to study the impact of such behaviour on the
system in more detail.

\subsection{Robustness against Random, Selfish, and Malicious Agents}

Another extension of the model focuses on the robustness of the
recommendation system against attacks. For this purpose, three different
additional types of agents can be considered: (1) \textit{Random agents}
are agents that, instead of giving correct recommendations, give a random
recommendation. Having such agents in the system mimics the effect of
noise on communication channels. (2) \textit{Selfish agents} are agents
that do not return recommendations except in the case that they have
already received responses through the agent that initiated the query.
(3) \textit{Malicious agents} are agents that intentionally give
recommendations that do not correspond to their own beliefs -- i.e., they
recommend what they would not use themselves, and vice versa. We are
interested in the performance of the recommendation system with respect
to differing fractions of such agents in the system: To what extent is
the performance affected? Is there a critical value of the fraction of
such agents for which the recommendation system becomes unusable?  For
applications in reality, an analysis of these topics is crucial.

\section{Summary and Conclusions}

We have outlined a model for a trust-based recommendation system that
combines the concepts of social networking and trust relationships:
agents use their trust relationships to filter the information that they
have to process and their social network to reach knowledge that is
located far from them. Probably the most striking result of this work is
that the recommendation system self-organises in a state with performance
near to the optimum; the performance on the global level is an emergent
property of the system, achieved without explicit coordination from the
local interactions of agents. With this model, we strive towards building
an archetypal model for recommendation systems by combining the
concepts of social networking and trust relationships.

\nocite{battiston-walter06}

\linespread{0.88} \normalsize

\bibliographystyle{acm}
\bibliography{paper}

\end{document}